\documentclass[runningheads]{llncs}
\usepackage[T1]{fontenc}

\usepackage{diagbox}
\usepackage{graphicx}
\usepackage{graphicx}
\usepackage{xcolor}
\usepackage{makeidx}  
\usepackage[colorlinks,linkcolor=blue]{hyperref}
\usepackage{amsmath, amsfonts}
\usepackage{booktabs}
\usepackage[misc]{ifsym} 
\usepackage{bbding}
\newcommand{\repeatthanks}{\textsuperscript{\thefootnote}}
\begin{document}

%
\title{FFPN: Fourier Feature Pyramid Network for Ultrasound Image Segmentation}
\titlerunning{FFPN}
%

\author{Chaoyu Chen\inst{1,2,3}\thanks{Chaoyu Chen and Xin Yang contribute equally to this work.} \and Xin Yang\inst{1,2,3}\repeatthanks \and Rusi Chen\inst{1,2,3} \and Junxuan Yu\inst{1,2,3} \and Liwei Du\inst{1,2,3} \and Jian Wang\inst{4} \and Xindi Hu\inst{5} \and Yan Cao\inst{5} \and Yingying Liu\inst{6} \and Dong Ni\inst{1,2,3}\textsuperscript{(\Letter)}} 

\institute{
\textsuperscript{$1$}National-Regional Key Technology Engineering Laboratory for Medical Ultrasound, School of Biomedical Engineering, Health Science Center, Shenzhen University, China\\
\email{nidong@szu.edu.cn} \\
\textsuperscript{$2$}Medical Ultrasound Image Computing (MUSIC) Lab, Shenzhen University, China\\
\textsuperscript{$3$}Marshall Laboratory of Biomedical Engineering, Shenzhen University, China\\
\textsuperscript{$4$}School of Biomedical Engineering and Informatics, Nanjing Medical University, China \\
\textsuperscript{$5$}Shenzhen RayShape Medical Technology Co., Ltd, China\\
\textsuperscript{$6$}Shenzhen People's Hospital, Second Clinical Medical College of Jinan University, China}

\authorrunning{C.Chen et al.}
%

%
\maketitle              
\begin{abstract}
Ultrasound (US) image segmentation is an active research area that requires real-time and highly accurate analysis in many scenarios. The detect-to-segment (DTS) frameworks have been recently proposed to balance accuracy and efficiency. However, existing approaches may suffer from inadequate contour encoding or fail to effectively leverage the encoded results. In this paper, we introduce a novel Fourier-anchor-based DTS framework called Fourier Feature Pyramid Network (FFPN) to address the aforementioned issues. The contributions of this paper are two fold. First, the FFPN utilizes Fourier Descriptors to adequately encode contours. Specifically, it maps Fourier series with similar amplitudes and frequencies into the same layer of the feature map, thereby effectively utilizing the encoded Fourier information. Second, we propose a Contour Sampling Refinement (CSR) module based on the contour proposals and refined features produced by the FFPN. This module extracts rich features around the predicted contours to further capture detailed information and refine the contours. Extensive experimental results on three large and challenging datasets demonstrate that our method outperforms other DTS methods in terms of accuracy and efficiency. Furthermore, our framework can generalize well to other detection or segmentation tasks.
\end{abstract}
\section{Introduction}
\begin{figure}
\centering
\scriptsize
\includegraphics[width=0.92\textwidth]{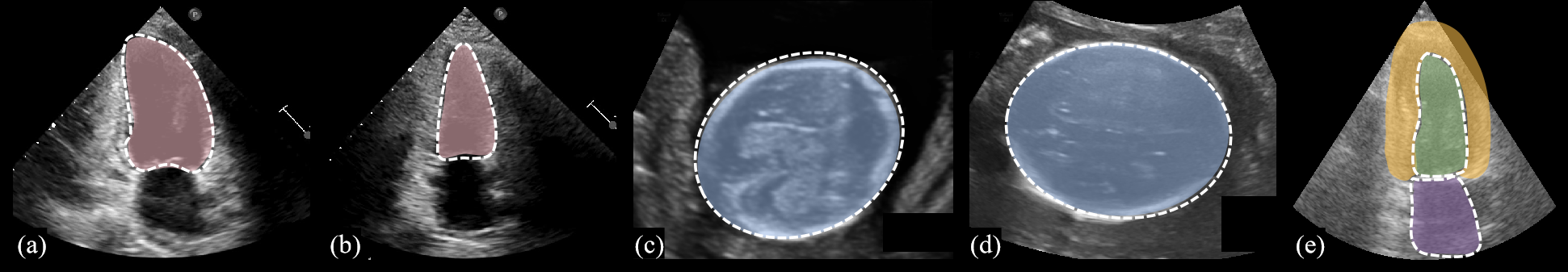}
\caption{Segmentation tasks in this paper. (a)-(b): 2CH data. (c)-(d): FH images. (e): Camus data.} \label{fig:task}
\end{figure}

Recently, real-time, accurate and low-resource image segmentation methods have gained wide attention in the field of ultrasound (US) image analysis. These methods provide the basis for many clinical tasks, e.g. structure recognition~\cite{camus_use}, bio-metric measurement~\cite{sobhaninia2019fetal} and surgical navigation~\cite{surgical_nav}. Fig.~\ref{fig:task} illustrates the segmentation tasks we have accomplished in this paper, including the apical two-chambers heart(2CH) dataset, Camus dataset~\cite{camus_data} and Fetal Head (FH) dataset.\par

Numerous segmentation methods based on deep learning have been proposed, most of which mainly rely on U-shaped networks, such as U-Net~\cite{ronneberger2015u}, nnU-Net~\cite{isensee2021nnu}, and SwinU-Net~\cite{cao2023swin}. The excessive skip-connections and upsampling operations in these methods make the model sacrifices efficiency and resources to ensure accuracy. DeeplabV3~\cite{chen2017rethinking} is another commonly used segmentation framework, and it also sacrifices efficiency due to the design of multiscale embedding. In addition, these methods all face the issue of false positive segmentation due to the blurred boundaries in US images. Thus, Mask R-CNN~\cite{he2017mask} uses the bounding box (b-box) as a constraint on the segmentation region, which reduces the false positive. However, the serial scheme limits its efficiency and excessive reliance on the b-box's output also affects its segmentation performance. \par

\begin{figure}
\centering
\scriptsize
\includegraphics[width=0.95\textwidth]{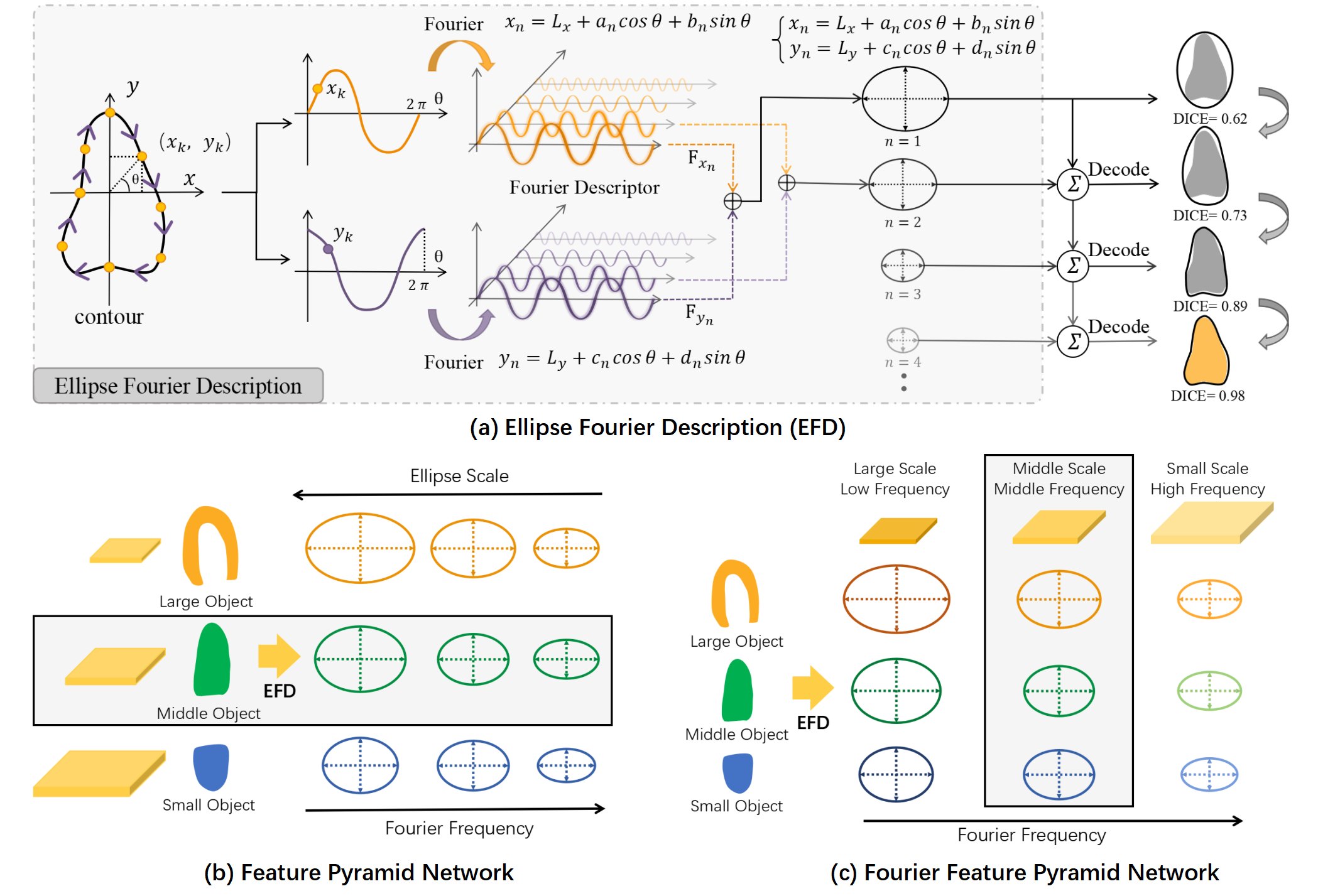}
\caption{(a) Detailed illustration of EFD scheme. (b) Previous methods only consider the scale and ignore the frequency characteristics. (c) Our FFPN focus on both the scale and frequency between different levels of Fourier series.}
\label{fig:EFD_FFPN_FIG}
\end{figure}

In order to balance resource consumption, efficiency and performance, detect-to-segment (DTS) framework has received significant attention in recent years.
The core idea of DTS is to transform the pixel classification problem of an image into the regression and classification problem for each point on the feature map. By this way, each point on the feature map can predict multiple contour proposals. PolarMask~\cite{xie2020polarmask} used a polar coordinate system for image segmentation, and it is difficult to process the situation that a ray intersects the object multiple times at a special direction. Point-Set~\cite{pointset} uniformly sampled several ordered points to represent the contour. PolySnake~\cite{polysnake} designed a multiscale contour refinement module to refine the initial coarse contour. While these sample-based encoding methods are capable of representing contours, many of the points within them do not contain valid information and there is a lack of correlation between points. Consequently, these encoding methods may not be suitable for segmentation tasks. To enhance the encoding results of complex contours, Ellipse Fourier Description~\cite{EFD} (EFD) scheme (Fig.~\ref{fig:EFD_FFPN_FIG} (a)) entered the vision of many researchers~\cite{fourier_only_21,fourier_22,CPN}. FCENet~\cite{fourier_only_21} and FANet~\cite{fourier_22} represent the text instance in the Fourier domain, allowing for fast and accurate representation of complex contours. CPN~\cite{CPN} revisited the EFD scheme to represent cell instance segmentation and achieved promising results. Although these EFD-based methods achieves high performance comparable to Pixel-based methods, they only consider the scale characteristics but ignore the frequency characteristics of the Fourier expansion, as shown in Fig.~\ref{fig:EFD_FFPN_FIG} (b), resulting in sub-optimal performance. \par

In this study, we revisited the EFD from another perspective, as shown in Fig.~\ref{fig:EFD_FFPN_FIG} (b) and (c). When the closed contour is expanded with Fourier, we focus on both the scale and frequency among different levels of Fourier series, and ingeniously incorporating it into the FPN~\cite{FPN} to devise a novel Fourier-anchor-based framework, named Fourier Feature Pyramid Network (FFPN). Our contributions are two fold. \textit{First}, we design FFPN to assign Fourier series with similar informcation to the same feature map for collective learning (Fig.~\ref{fig:EFD_FFPN_FIG} (c)). This approach enhances the consistency of feature representation and improves the model's ability to predict encoded results with better accuracy. \textit{Second}, considering the complexity and blurring of object contours in US images, we propose a Contour Sampling Refinement (CSR) module to further improve the model's ability to fit them. Specifically, we aggregate features at different scales in FFPN and extracted the relevant features around the contours on the feature map to rectify the original contour. Experimental results demonstrate that FFPN can stably outperform DTS competitors. Furthermore, our proposed FFPN is promising to generalize to more detection or segmentation tasks. \par
 
\section{Methodology}
\begin{figure}
\centering
\scriptsize
\includegraphics[width=0.95\textwidth]{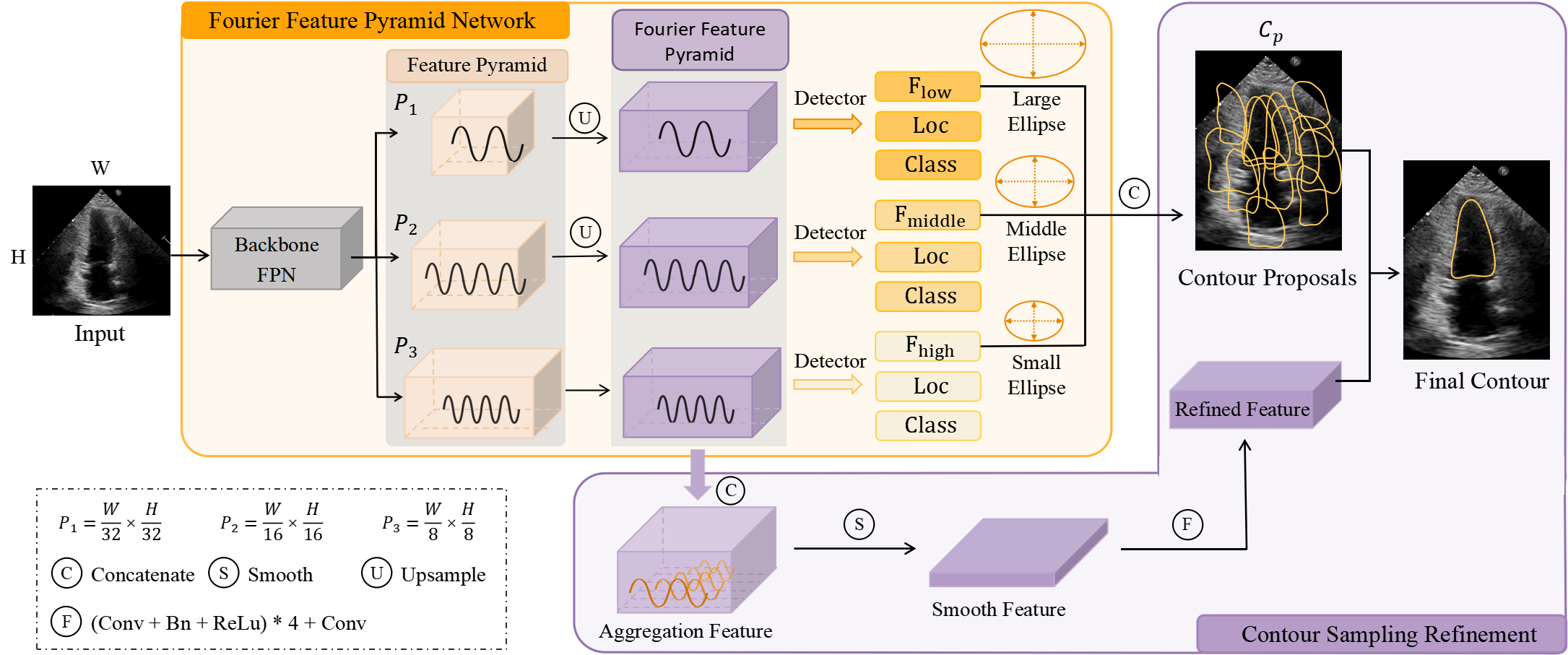} 
\caption{The overall pipeline of our proposed FFPN. In yellow blocks, $F$ denotes Fourier serie offset, $Loc$ denotes location offset, $Class$ denotes classification score.} 
\label{fig:framework}
\end{figure}

Fig.~\ref{fig:framework} is the overview of our FFPN framework. Given an image, we first use the backbone with FPN to extract the pyramidal features. Then these features are unsampled to obtain Fourier pyramidal features. Next, we feed Fourier pyramidal features into different detector heads to generate different levels of Fourier series offsets, location offsets and classification scores. EFD decodes these predictions to the contour proposals. In CSR, the Fourier pyramid features are concatenated to generate the refinement features. Finally, the contour proposals and refinement features together are used to produce the final results. \par

\subsection{Fourier Feature Pyramid Network(FFPN)}
Inspired by EFD (Fig.~\ref{fig:EFD_FFPN_FIG} (a)), the contour can be described as the Fourier series:
\begin{equation}
\centering
\footnotesize
\begin{array}{cc}
x_N(t) = L_x+\sum_{n=1}^{N}(a_nsin(\frac{2n\pi t}{T})+b_ncos(\frac{2n \pi t}{T} )), \\
\\
y_N(t) = L_y+\sum_{n=1}^{N}(c_nsin(\frac{2n\pi t}{T})+d_ncos(\frac{2n \pi t}{T})),
\end{array}
\label{equ:efd}
\end{equation}
where \begin{math}(x_N(t),y_N(t))\end{math} is the \begin{math}t\end{math}-th sampled point on the contour, and the number of sampled points are set to \begin{math}T\end{math}, e.g. \begin{math} t\in[0,T] \end{math}. \begin{math} N \end{math} is the number of Fourier series expansions. \begin{math}(L_x,L_y)\end{math} indicates the coordinates of the center point of the contour. \begin{math}a_n, b_n\end{math} denote the parameters obtained by Fourier coding of the x-coordinates of all contour points, and \begin{math}c_n, d_n\end{math} denote the corresponding parameters in the y-coordinates. Thus, the goal of FFPN is to accurately predict the \begin{math}4N+2\end{math} parameters to represent the contour (\begin{math}N = 7\end{math} by default).\par

According to EFD (Fig.~\ref{fig:EFD_FFPN_FIG} (a)), the low-level Fourier series represent the contour's low-frequency information and main scales, while the high-level Fourier series capture the contour's high-frequency information and shape details. This is consistent with the extracted pyramidal features, where the low-level features \begin{math}(P_3)\end{math} contain more detailed information, while the high-level features \begin{math}(P_1)\end{math} capture semantics. Therefore, the proposed FFPN effectively aggregates Fourier series of similar scales and frequencies into the same feature map (as illustrated in Fig. ~\ref{fig:EFD_FFPN_FIG} (c)). Then, different level features of feature pyramid are fed into different detector heads with the same architecture (three sibling 3x3 Conv-BN-ReLUs followed by a 3x3 Conv) to generate Fourier offsets, location offsets and classification scores ($F$, $Loc$ and $Class$ with yellow blocks in Fig.~\ref{fig:framework}.). The up-sampling operations on the \begin{math}P_1\end{math} and \begin{math}P_2\end{math} features are only to align the scale and simplify subsequent operations. Next, the different level Fourier offsets are concatenated along the channel dimension to predict contour proposals. To simplify the calculation, location offsets and classification scores at different levels are averaged along the channel dimension. The learning objectives are as follows:

 
\begin{equation}
\centering
\footnotesize
    \left\{
        \begin{array}{ll}
            \Delta {F_{a}}_{i} = \frac{{G_{a}}_{i}-{A_{a}}_{i}}{{E_{x}}_{i}} \\
            \\
            \Delta {F_{b}}_{i} = \frac{{G_{b}}_{i}-{A_{b}}_{i}}{{E_{x}}_{i}} \\
        \end{array}
    \right.\;
    \hspace{3em}\left\{
        \begin{array}{ll}
            \Delta {F_{c}}_{i} = \frac{{G_{c}}_{i}-{A_{c}}_{i}}{{E_{y}}_{i}} \\
            \\
            \Delta {F_{d}}_{i} = \frac{{G_{d}}_{i}-{A_{d}}_{i}}{{E_{y}}_{i}} \\
        \end{array}
    \right.
    \hspace{3em}\left\{
        \begin{array}{ll}
            \Delta L_x = \frac{{G_{L}}_{x}-{A_{L}}_{x}}{{E_{x}}_{1}}  \\
            \\
            \Delta L_y = \frac{{G_{L}}_{y}-{A_{L}}_{y}}{{E_{y}}_{1}},  \\
        \end{array}
    \right.
\label{equ:encode}
\end{equation}
where the tuple \begin{math}({G_{a}}_{i}, {G_{b}}_{i}, {G_{c}}_{i}, {G_{d}}_{i})\end{math} represents the Ground truth(GT) of the Fourier series, and \begin{math}({A_{a}}_{i}, {A_{b}}_{i}, {A_{c}}_{i}, {A_{d}}_{i})\end{math} denotes the Fourier series of anchor. \begin{math}i\in[1,N]\end{math} denotes the \begin{math}i\end{math}-th of the Fourier expansion. The Fourier expansion of each level can be expressed as an ellipse, as shown in Fig.~\ref{fig:EFD_FFPN_FIG} (a).
\begin{math}{E_{x}}_{i}\end{math} and \begin{math}{E_{y}}_{i}\end{math} denote the width and height of the ellipse at level \begin{math}i\end{math}, respectively.
\begin{math}({G_{L}}_{x}, {G_{L}}_{y})\end{math} is the center point coordinates of the GT contour, and \begin{math}({A_{L}}_{x}, {A_{L}}_{y})\end{math} is the center point coordinates of the anchor. \begin{math}(\Delta {F_{a}}_{i}, \Delta {F_{b}}_{i}, \Delta {F_{c}}_{i}, \Delta {F_{d}}_{i}, \Delta L_x, \Delta L_y)\end{math} are the learning objectives. Through the encoding operation of Eq.~\ref{equ:encode}, we normalize the amplitude of Fourier series and center point coordinates to the same scale, simplifying the learning difficulty. The loss function is defined as follows:
\begin{equation}
\centering
\footnotesize
{\mathcal{L}} = {\mathcal{L}}_{\tiny Loc}+{\mathcal{L}}_{\tiny Fou}+{\mathcal{L}}_{\tiny Con}+ {\mathcal{L}}_{\tiny Cls}
\label{equ:loss_func}
\end{equation}
The loss function about Location and Fourier are Smooth L1 Loss, and the classification loss is \begin{math}{\mathcal{L}}_{\tiny Cls} = \alpha BCELoss + \beta FocalLoss\end{math}, where \begin{math}\alpha = 0.25\end{math} and \begin{math}\beta = 0.75\end{math}. The contour loss is defined as \begin{math}{\mathcal{L}}_{\tiny Con} = 1 - PolarIoU*BoxIoU\end{math}, where the \begin{math} PolarIoU \end{math} follows~\cite{xie2020polarmask} and the \begin{math} BoxIoU \end{math} is defined as the IoU between outline bounding boxes. The method to calculate IoU in this paper is defined as \begin{math} IoU = PolarIoU*BoxIoU\end{math}, which is a simple but effective way.

\subsection{Contour Sampling Refinement (CSR) Module}
Considering the complexity of contour representation, we adopt the two-stage strategy inspired from the detection frameworks~\cite{FPN,he2017mask}. Unlike CPN~\cite{CPN} where the contour is directly optimized to destroy its smoothness or FANet~\cite{fourier_22} where the Fourier series of the contour needs to be refined iteratively, we refine once the Fourier series of the aggregation averaged contour. This strategy improves the model accuracy while ensures the boundary smoothness. Specifically, as shown in Fig.~\ref{fig:framework}, we fuse pyramidal features as the refined-feature. And the top-n \begin{math}(n = 20)\end{math} outputs of FFPN are the contour proposals \begin{math}(C_p)\end{math} to represent the same object. Thus, the refined-feature and the \begin{math}C_p\end{math} are the input of CSR module. \par
\begin{figure}
\centering
\scriptsize
\includegraphics[width=0.95\textwidth]{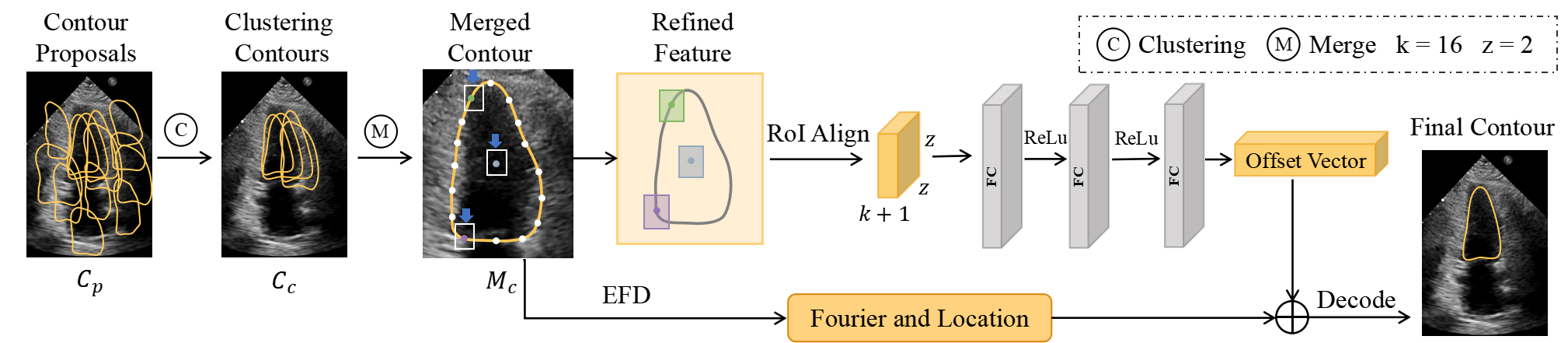}
\caption{Illustration of our proposed Contour Sampling Refinement (CSR) module.}
\label{fig:refine_moduel}
\end{figure}
Fig.~\ref{fig:refine_moduel} demonstrates the workflow of CSR. \textit{First}, 
we extract the closely clustered contours \begin{math}(C_c)\end{math} in \begin{math}C_p\end{math}. The definition of closely clustered is that: at least one contour exists, and the IoU of this contour with all the remaining contours is greater than a threshold \begin{math}(t = 0.7)\end{math}. \textit{Second}, we obtain the merged contour \begin{math}(M_c)\end{math} of the object by averaging the \begin{math}C_c\end{math} directly. We think that, as the network learns iteratively, \begin{math}M_c\end{math} has a high confidence in fitting the object. Thus, we extract the feature around \begin{math}M_c\end{math} to further refine \begin{math}M_c\end{math}. Concretely, we use the uniform point sampling approach to sample out \begin{math}(k+1)\end{math} sampling points, indicating the \begin{math}k\end{math} boundary points and one center point of the contour\begin{math}(k=16)\end{math}. To extract the features around the contour, we use the sampled points as center points to generate \begin{math}(k+1)\end{math} boxes to obtain rich information on refined-feature. RoI-Align module is used in this stage. Finally, these extracted features are passed through a three-layer Multi-Perceptron module to generate the Fourier and center point offsets. The losses used in CSR are same as FFPN, including \begin{math}{L}_{\tiny Fou}\end{math}, \begin{math}{L}_{\tiny Loc}\end{math} and \begin{math}{L}_{\tiny Con}\end{math}. It is worth noting that the anchor information used at this stage is generated by \begin{math}M_c\end{math}, which draws on the common configuration of a two-stage detection framework.

\section{Experimental Results}
To validate the performance of our FFPN, we conducted comprehensive comparisons among our method and other segmentation methods, including DTS methods and Pixel-based methods. Additionally, we also compare FFPN with FFPN$+$CSR (FFPN-R) to validate the effectiveness of CSR. 

\textbf{Datasets.} We assess FFPN's performance across three datasets (Fig.~\ref{fig:task}). Approved by local IRB, the 2CH dataset, comprising 1731 US images, is utilized for left ventricle (LV) segmentation, while the FH dataset, containing 2679 US images, is employed for fetal head segmentation. The Camus~\cite{camus_data} dataset contains 700 US images and requires the segmentation of three structures, including the LV, left atrium (LA) and myocardium (MC). Both the 2CH dataset and the FH dataset have been manually annotated by experts using the Pair~\cite{pair} annotation software package. Each dataset undergoes a random split into training (70\%), validation (10\%), and testing (20\%) subsets.\par

\textbf{Experimental Settings.} To conduct a fair comparison, all methods implemented in Pytorch and under the same experiment settings, including learning rate (1e-3), input size (directly resized to 416x416), total epochs (200), one RTX 2080Ti GPU and so on. We evaluate the model performance using dice similarity coefficient (DICE), Hausdorff distance (HD), Intersection over Union (IoU) and Conformity (Conf) for all the experiments. Memory (Mem) and FPS are employed to evaluate the models' memory usage and efficiency. As for the settings of FFPN, the Fourier-anchors are generated by contour clustering of the training set, each dataset has 9 base anchors. The IoU threshold for positive and negative samples are 0.25 and 0.10 respectively, and the others are ignored samples. \par

\begin{table}[]
        \renewcommand{\arraystretch}{1.25}
        \scriptsize
        \centering
        \caption{Quantitative evaluation of mean(std) results on 2CH and FH datasets.}
        \resizebox{\linewidth}{!}{
        \setlength{\baselineskip}{5em}{
    
        \begin{tabular}{c|cccc|ccl|cc}
        \hline
        DATASET         & \multicolumn{4}{c|}{2CH}                      & \multicolumn{3}{c|}{FH}                                  &\multicolumn{2}{c}{Model}       \\ \hline
        \diagbox{Methods}{Metrics}        & DICE($\%$)$\uparrow$   & IoU($\%$)$\uparrow$     & HD(pixel)$\downarrow$    & Conf($\%$)$\uparrow$      & DICE($\%$)$\uparrow$        & HD(pixel)$\downarrow$          & Conf($\%$)$\uparrow$   & Mem(G)$\downarrow$  & FPS$\uparrow$   \\ \hline
        U-NET           & 88.95(4.95)     & 80.45(7.55)      & 22.41(18.80)        & 74.40(13.95)    & 94.92(3.27)     & 18.1(16.43)       & 90.78(6.87)   & 1.33 & 19.71 \\
        DeepLabV3       & 88.92(5.06)     & 80.40(7.69)      & 19.98(12.14)        & 74.26(14.55)    & 95.89(2.56)     & 16.63(22.59)       & 91.27(6.33)    & 0.49 & 15.83  \\
        Swin U-NET      & 88.83(4.63)     & 80.21(7.21)      & 20.25(10.28)        & 74.20(12.67)    & 95.81(1.38)     & 15.30(5.67)        & 91.21(3.10)    & 2.27 & 9.81  \\
        Mask RCNN       & 79.50(6.71)     & 66.48(9.04)      & 47.53(13.87)        & 46.51(22.90)    & 82.60(2.33)     & 58.24(6.73)        & 57.64(7.10)    & 0.43  & 20.33  \\\hline
        PolarMask       & 84.54(14.65)    & 77.42(13.15)     & 34.13(25.81)        & 58.96(78.22)    & 94.25(5.31)     & 25.96(15.12)       & 64.28(489.90)  & 0.52  & 15.43  \\
        PolySnake       & 86.50(6.05)     & 76.67(8.73)      & 22.75(13.05)        & 67.48(18.82)    & 92.11(4.39)     & 22.95(11.46)       & 82.32(11.43)   & \textcolor{blue}{0.17}  & 14.73  \\
        CPN             & 87.62(6.23)     & 78.47(9.02)      & 23.81(15.81)        & 72.61(15.30)    & 94.00(5.93)     & 22.73(22.89)       & 86.17(16.83)   & 0.20  & 27.20  \\
        FFPN            & 88.16(5.97)     & 79.30(8.88)      & 21.10(14.34)        & 71.93(18.15)    & 95.56(2.40)     & 13.90(8.84)        & 90.55(6.56)  & 0.20  & \textcolor{blue}{41.52}  \\
        \textbf{FFPN-R} &  \textcolor{blue}{89.08(5.24)}  & \textcolor{blue}{80.70(8.10)}   &  \textcolor{blue}{19.76(12.52)} &  \textcolor{blue}{74.64(14.55)}   &  \textcolor{blue}{96.73(1.11)}  &  \textcolor{blue}{10.24(4.13)}   &  \textcolor{blue}{93.21(2.43)}   & 0.23  & 33.52  \\ \hline
        \end{tabular}
    }
}
\label{main_result1}
\end{table}

\textbf{Quantitative and Qualitative Analysis.} 
As demonstrated in Table~\ref{main_result1}, FFPN exhibits superior performance compared to feasible DTS methods and achieves comparable performance when compared to effective Pixel-based methods in both the 2CH dataset and FH dataset. Specifically, comparing to the recently proposed method PolySnake, FFPN increases DICE by 1.66\% and 3.45\%, while simultaneously reducing HD by 1.65 pixels and 9.05 pixels on the 2CH dataset and FH dataset, respectively. This demonstrates the effectiveness of our encoding approach based on EFD. Moreover, in comparison to CPN, which follows the same encoding approach, FFPN improves DICE by 0.54\% and 1.56\%, and decreases HD by 2.71 pixels and 8.83 pixels on the two datasets, respectively. This fully proves that FFPN effectively utilizes Fourier information by mapping Fourier series with similar scales and frequencies into the same layer of the feature map. Furthermore, FFPN-R is capable of outperforming all methods in all metrics on the two datasets. This demonstrates the effectiveness of CSR, particularly in its ability to capture detailed information and refine contours.\par
Experimental results on computing resource consumption and efficiency are shown in the \begin{math} Model \end{math} column of Table~\ref{main_result1}. FFPN achieves an impressive inference speed of 41.52 FPS with a memory consumption of only 0.2 GB, making it the fastest among all DTS and Pixel-based methods. Compared with PolySnake which has the smallest memory footprint, FFPN increases memory usage by only 17.6\% while improving its speed by 182.0\%. Furthermore, by incorporating CSR, FFPN-R achieves the best accuracy performance at a slight cost of increasing memory consumption by 15.0\% and reducing inference speed by 19.5\% FPS. Despite this minor trade-off, FFPN-R still exhibits superior resource consumption and efficiency compared to all Pixel-based and most DTS methods. \par
\begin{figure}
\centering
\scriptsize
\includegraphics[width=1\textwidth]{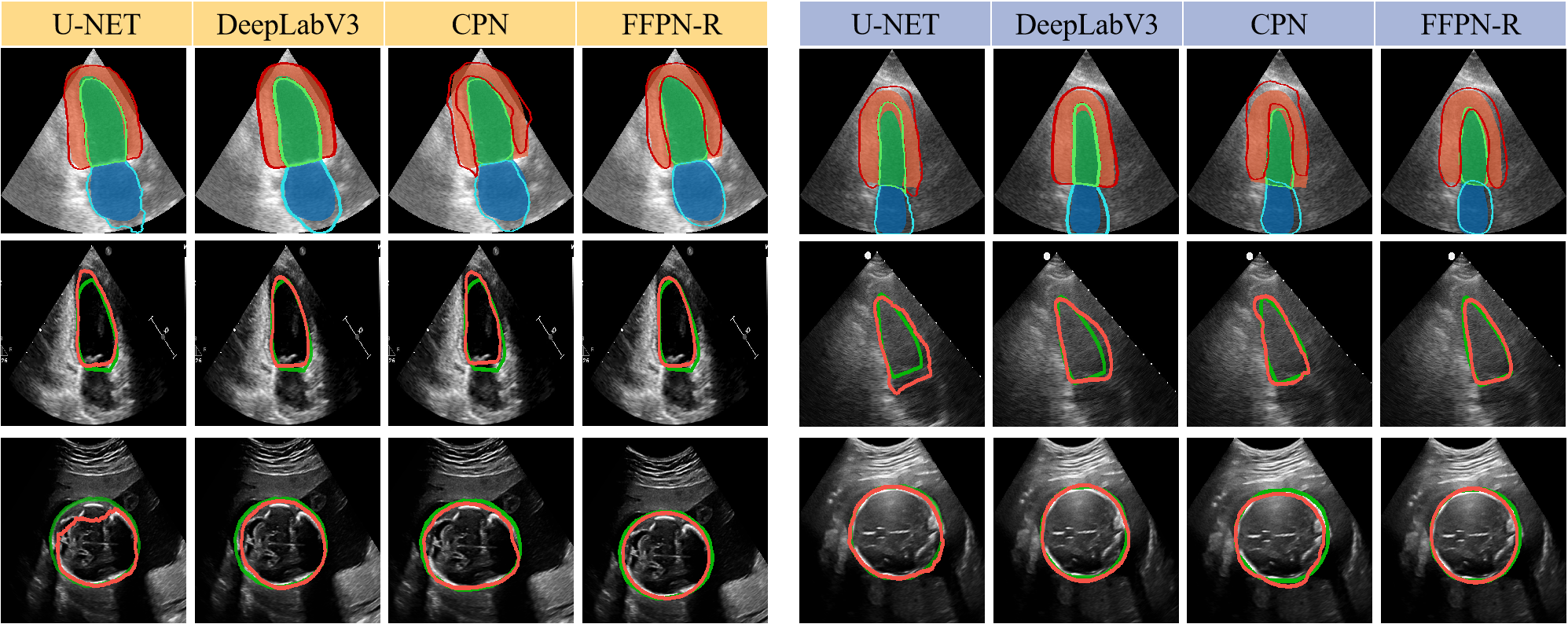}
\caption{Segmentation results of different methods on three datasets. For the first row (Camus), the masks are ground truth, and the contours are predictions. For the second (2CH) and third row (FH), the green contours denote ground truth, and the red contours represent predictions.}
\label{fig:img_compare}
\end{figure}

\begin{table}[]
            \renewcommand{\arraystretch}{1.25}
            \scriptsize
            \centering
            \caption{Quantitative evaluation of mean(std) results on Camus dataset.}
            \resizebox{\linewidth}{!}{
            \setlength{\baselineskip}{5em}{
            \begin{tabular}{c|cc|cc|cc|cc}
            \hline
                       & \multicolumn{2}{c|}{LA}     & \multicolumn{2}{c|}{MC}     & \multicolumn{2}{c|}{LV}     & \multicolumn{2}{c}{Mean}    \\ \hline
           \diagbox{Methods}{Metrics}    & DICE($\%$)$\uparrow$     & HD(pixel)$\downarrow$           & DICE($\%$)$\uparrow$     & HD(pixel)$\downarrow$           & DICE($\%$)$\uparrow$     & HD(pixel)$\downarrow$           & DICE($\%$)$\uparrow$    & HD(pixel)$\downarrow$            \\ \hline
            U-NET      & 87.74(9.81)          & 28.88(32.99)          & 86.03(4.45)          & 22.43(21.84)          & 92.06(4.01)          & 22.43(28.94)         & 88.61(6.09)          & 24.58(27.78)          \\
            DeepLabV3  & \textcolor{blue}{88.93(6.34)} & \textcolor{blue}{22.84(21.91)}          & \textcolor{blue}{87.84(4.04)} &\textcolor{blue}{17.49(13.05)} & \textcolor{blue}{92.79(4.02)} & \textcolor{blue}{16.51(9.36)} & \textcolor{blue}{89.85(4.80)} & \textcolor{blue}{18.69(18.80)} \\
            Swin U-NET & 87.28(9.47)          & 26.39(24.97)          & 86.38(4.32)          & 19.31(8.60)           & 91.73(4.71)          & 18.68(11.69)         & 88.46(6.17)          & 21.46(15.10)          \\
            Mask RCNN  & 76.83(18.61)         & 43.38(40.57)          & 55.99(26.54)         & 66.78(84.70)          & 75.16(16.99)         & 53.95(32.10)         & 69.33(20.71)         & 54.70(52.46)          \\ \hline
            PolarMask  & 81.56(20.44)         & 40.91(16.25)          & 27.28(21.16)         & 59.92(12.98)          & 81.56(20.44)         & 42.67(37.64)         & 65.75(16.23)         & 47.83(22.29)          \\
            PolySnake  & 83.77(17.23)         & 22.99(19.92)          & 43.65(32.27)         & 21.82(11.43)          & 89.58(9.80)          & 19.72(11.30)         & 72.33(19.77)         & 21.51(14.21)          \\
            CPN        & 87.11(13.68)         & 27.35(37.09)          & 69.17(25.75)         & 41.58(58.22)          & 92.08(5.06)          & 19.26(13.62)         & 82.79(14.83)         & 29.40(36.31)          \\
            FFPN       & 87.37(8.39)          & 23.81(17.18)          & 82.52(6.59)          & 23.63(9.83)           & 91.35(4.36)          & 20.02(11.12)         & 87.08(7.57)          & 22.48(13.23)          \\
            \textbf{FFPN-R}     & \textcolor{blue}{88.76(7.85)}          & \textcolor{blue}{21.10(17.65)} & \textcolor{blue}{85.03(4.54)}          & \textcolor{blue}{20.25(7.43)}           & \textcolor{blue}{92.39(4.08)}          & \textcolor{blue}{16.97(9.36)}         & \textcolor{blue}{88.72(6.48)}          & \textcolor{blue}{19.44(12.43)}          \\ \hline
            \end{tabular}
            }
            }
    \label{main_result2}
\end{table}
In addition, we have also validated the effectiveness of our framework in the multi-class segmentation task. As shown in Table~\ref{main_result2}, FFPN-R achieves a more significant improvement on the Camus dataset compared to other DTS approaches, as compared to the single-class segmentation tasks. This further illustrates the generalizability and scalability of our framework. On the Camus dataset, FFPN-R outperforms most Pixel-based methods. However, the wrapping of the myocardium around the left ventricle poses a challenge in accurately assigning positive and negative samples within the detection framework, resulting in our results being slightly inferior to DeeplabV3. Fig.~\ref{fig:img_compare} shows the segmentation results of U-NET, DeepLab V3, CPN and FFPN-R. It demonstrates the superior performance of FFPN-R on the segmentation task in US images.\par

\section{Conclusion}
In this study, we utilize the effectiveness of Fourier Descriptors to represent contours and propose Fourier Feature Pyramid Network (FFPN), a Fourier-anchor-based framework, to describe the segmentation region. It is found that FFPN  achieves the best performance against other DTS methods. Moreover, we design a Contour Sampling Refinement (CSR) module to more accurately fit complex contours, which makes our method achieve further improvement and exhibit powerful capabilities on segmentation tasks. These experiments further demonstrate the well-balanced among performance, resource consumption and efficiency of our framework.
\subsubsection{Acknowledge.}
This work was supported by the grant from National Natural Science Foundation of China (Nos. 62171290, 62101343), Shenzhen-Hong Kong Joint Research Program (No. SGDX20201103095613036), and Shenzhen Science and Technology Innovations Committee (No. 20200812143441001).

%
%
%
%

\bibliographystyle{splncs04}
\bibliography{ref.bib}

\end{document}